\documentclass[12pt]{iopart}
\usepackage{iopams}
\usepackage{graphicx}

\begin{document}

\title[Sedimentation of colloidal gels]{Equilibrium concentration profiles and sedimentation kinetics of colloidal gels under gravitational stress}

\author{S. Buzzaccaro$^1$, E. Secchi$^1$,  G. Brambilla$^{2,3,4}$, R. Piazza$^1$, L. Cipelletti$^{2,3}$}

\address{$^1$Dipartimento di Chimica, Politecnico di Milano, 20131
Milano, Italy\\
$^2$Universit\'{e} Montpellier 2, Laboratoire Charles Coulomb UMR 5221, F-34095, Montpellier,
France\\
$^3$CNRS, Laboratoire Charles Coulomb UMR 5221, F-34095,
Montpellier, France $^4$ Present address: Formulaction, L'Union,
France. }
\ead{stefano.buzzaccaro@mail.polimi.it,luca.cipelletti@univ-montp2.fr}
\begin{abstract}
We study the sedimentation of colloidal gels by using a combination
of light scattering, polarimetry and video imaging. The asymptotic
concentration profiles $\varphi(z,t\rightarrow \infty)$ exhibit
remarkable scaling properties: profiles for gels prepared at
different initial volume fractions and particle interactions can be
superimposed onto a single master curve by using suitable reduced
variables. We show theoretically that this behavior stems from a
power law dependence of the compressive elastic modulus \textit{vs}
$\varphi$, which we directly test experimentally. The sedimentation
kinetics comprises an initial latency stage, followed by a rapid
collapse where the gel height $h$ decreases at constant velocity,
and a final compaction stage characterized by a stretched
exponential relaxation of $h$ towards a plateau. Analogies and
differences with previous works are briefly discussed.

\end{abstract}

\submitto{\JPCM}
%\maketitle

\section{Introduction}
Colloidal gels are the focus of an intensive research effort for
both fundamental and practical reasons. On the one hand, they are
model systems to understand the interplay between percolation, phase
separation and dynamical arrest in systems with attractive
interactions~\cite{ZaccarelliJPCM2007,Buzzaccaro,LuNature2008}. On
the other hand, colloidal gels are ubiquitous, e.g., in the food,
drug, personal care and cosmetic industries, where they are often
used as a means to stabilize a complex formulation against
macroscopic phase separation. Mechanically, colloidal gels are
viscoelastic systems with a predominantly solid-like behavior.
However, they typically yield under a modest stress, often including
the gravitational stress exerted by their own weight. It is
therefore not surprising that many studies have been devoted to the
sedimentation behavior of colloidal gels, revealing a wealth of
fascinating but yet not fully understood phenomena, such as delayed
sedimentation~\cite{StarrsJPCM2002,kilfoil03,huh07,buscall09,BartlettArxiv},
creep~\cite{StarrsJPCM2002,AllainPRL1995,CondreJSTAT2007,ManleyPRL2005Sedimentation,KimPRL2007}, and fracture
associated with complex flow
patterns~\cite{StarrsJPCM2002,AllainPRL1995,DerecPRE2003}. With a few notable
exceptions~\cite{StarrsJPCM2002,huh07,BartlettArxiv}, most previous work has
focussed on the macroscopic behavior of the gels, typically by
measuring the time evolution of the gel height, $h(t)$. This has
prevented a thorough test of the various models proposed to
rationalize the gel behavior, and in particular of the poroelastic
model~\cite{BiotJAP1941,BuscallFaradyTrans1987}.

To overcome these limitations, we have recently reported in
Ref.~\cite{BrambillaPRL2011} a full characterization of the temporal
evolution of the concentration profile, $\varphi(z,t)$,
sedimentation velocity profile, $v(z,t)$ and microscopic
rearrangement dynamics of a colloidal gel. In particular, we have
shown that the poroelastic model captures remarkably well the
evolution of $h(t)$, $\varphi(z,t)$ and $v(z,t)$. Additionally, a
single quantity, the compressive strain rate $\dot{\varepsilon}(t)$,
was shown to control both the macroscopic behavior and the
microscopic dynamics. Here, we test the generality of the findings
of Ref.~\cite{BrambillaPRL2011} by studying gels with different
particle concentration and interparticle attraction strength. We
focus on the sedimentation kinetics and on the asymptotic
concentration profiles as the gels attain mechanical equilibrium,
showing that the concentration profiles have a universal shape.
Using the poroelastic model~\cite{BiotJAP1941,BuscallFaradyTrans1987}, we show that this
remarkable scaling behavior stems from a particularly simple
relation between elastic stress and local volume fraction. The rest
of the paper is organized as follows: in Sec.~\ref{sec:m&m} we
describe the system and the optical methods used to investigate it.
The main features of the poroelastic model are recalled in
Sec.~\ref{sec:poroelastic}, while
Secs.~\ref{sec:equilibriumprofiles} and~\ref{sec:kynetics} report
our findings on the asymptotic concentration profiles and the
sedimentation kinetics, respectively, before the concluding remarks
of Sec.~\ref{sec:conclusions}.

\section{Materials and methods}
\label{sec:m&m}

\subsection{Sample preparation}

The colloidal particles are spheres of a polytetrafluoroethylene
copolymer (MFA). They have a crystalline core, leading to peculiar
optical properties that allow one to measure accurately the local
volume fraction, as discussed below. The data presented here are
obtained using two different batches, B1 and B2, for which the
particle radii are slightly different: $R_1=82 \pm 3$ nm and $R_2=92
\pm 3$ nm, for batches B1 and B2 respectively. The high density
$\rho = 2.14~\mathrm{g/cm^3}$ of MFA, combined with its lack of
swelling, allows for a precise determination of the initial particle
volume fraction $\varphi_0$ by density measurements. The particles
are suspended in an aqueous solution of 0.1M NaCl, to screen
electrostatic repulsive interactions, and different amounts of urea
were added to suppress coherent polarized scattering by matching the
solvent and particle refractive indices. A nonionic surfactant,
Triton X100, is added to induce attractive depletion
interactions~\cite{AsakuraJChemPhys1954}, whose range $r \approx 3$
nm is of the order of the size of the micelles formed by the
surfactant. In the following we denote by $\phi_{\rm TX}$ the
concentration of surfactant, defined by $\phi_{\rm TX} = V_{\rm
TX}/(V_s-V_p)$, where $V_{\rm TX}$, $V_p$ and $V_s$ are the volume
of the surfactant, that of the particles, and the total volume of
the suspension, respectively. A detailed description of the phase
behavior of this system can be found in~\cite{Buzzaccaro}. In brief,
the system displays a metastable liquid-liquid coexistence gap; when
the amount of added depletant is sufficiently large to drive the
colloidal suspension within the coexistence region, as in our case,
arrested spinodal decomposition leads to the formation of a
disordered gel. All samples described here are deeply in the
coexistence region, far from the spinodal line. Because the range of
the attractive potential is much smaller than the particle size, our
system is well described by Baxter's adhesive hard sphere
model~\cite{BaxterJChPh1968}. It is convenient to quantify the
strength of the attractive interactions via Baxter's stickiness
parameter $\tau_{\rm B}$. Using methods detailed
in~\cite{Buzzaccaro}, we estimate $\tau \approx 0.01$ for the gels
of batch B1 and $\tau_{\rm B} \approx 0.02$ for the gels of batch B2 at
$\Phi_{\rm TX} = 0.11$.
%This eases a direct comparison with other systems by virtue
%of the Noro-Frenkel law of corresponding states, which states that
%all short-ranged spherically symmetric pair-wise additive attractive
%potentials are characterized by the same thermodynamics properties
%if compared at the same reduced density and second virial
%coefficient~\cite{noro00}

The samples are prepared by mixing a particle suspension and a
surfactant solution with appropriate concentrations; the resulting
suspension is loaded in cells with square cross section (either $3
\times 3~\mathrm{mm}^2$ or $5 \times 5~\mathrm{mm}^2$) and filled to
an initial height $h_0 = 22.7~\mathrm{mm}$ ($h_0 = 32~\mathrm{mm}$)
for batch B1 (B2). The samples are shaken at time $t_{\rm w}= 0$ and
then left undisturbed for measurements during the sedimentation
process.

\subsection{Light scattering measurements}
\begin{figure}
\begin{center}
\includegraphics[width = 8 cm]{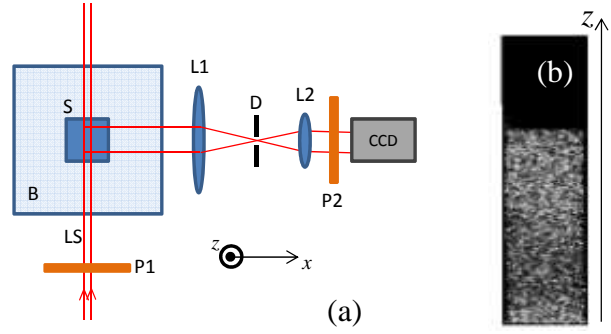}
\end{center}
\caption{(a): schematic top view of the light scattering apparatus.
See the text for more details. (b): Typical image of the sample. The
gel is the bright, speckled column, the black part is the
supernatant.} \label{fig:apparatus}
\end{figure}

The apparatus used for light scattering measurements on systems
prepared from the first batch is shown in Fig.~\ref{fig:apparatus}
(a). The sample S is immersed in a transparent water bath B to
minimize temperature gradients and temperature fluctuations. It is
illuminated by a laser sheet LS, of thickness $\approx 200~\mu{\rm
m}$, height sufficient to cover the whole gel column, and in-vacuo
wavelength $\lambda = 648~{\rm nm}$. The two lenses L1 and L2 form a
demagnified image of the sample onto a charge-coupled detector (CCD)
camera. Their focal length is 200 mm and 58 mm, respectively. Two
crossed polarizers, P1 and P2, are used to illuminate the sample
with light polarized in the vertical direction while detecting only
depolarized scattered light, i.e. light linearly polarized in the
horizontal direction. Note that this apparatus combines features of
both imaging and scattering~\cite{DuriPRL2009}, since an image of
the sample is formed, but using only light scattered around a
well-defined scattering wave vector $q = 4\pi n \lambda^{-1}
\sin(\theta/2) = 18.5~\mu{\rm m}^{-1}$, with $n = 1.356$ the refractive
index of the solvent and $\theta = 90$ deg the scattering angle. A
typical image recorded by the CCD is shown in
Fig.~\ref{fig:apparatus} (b). The gel appears as a speckled, bright
column, while the supernatant is black since scattering by the
solvent is negligible. Each speckle results from the interference of
the light scattered by a small volume in the sample, of depth equal
to the laser sheet thickness and lateral size $\approx 50 ~\mu{\rm
m}$. The speckle
size is controlled to about 2 CCD pixels by
adjusting the range of $\theta$ accepted by the detection optics,
using the diaphragm D placed in the common focal plane of L1 and L2.

By analyzing a time series of CCD images, we measure four different
quantities. The gel total height $h(t)$ is simply obtained from the
height of the bright column. The concentration profiles
$\varphi(z,t)$ are obtained by averaging the intensity $I$ over
regions of interest (ROIs) of height $\approx 100 ~\mu {\rm m}$ and
width equal to the cell width. We emphasize that in general the
(polarized) scattered intensity is proportional to both $\varphi$
and the particle structure factor $S(q)$, thus making impossible a
determination of the volume fraction based on $I$ without a detailed
knowledge of the sample structure. By contrast, our particles posses
a crystalline core and thus partially depolarize the scattered
light. As discussed, e.g., in Ref.~\cite{buzzaccaro3}, the
depolarized scattered light can be shown to be simply proportional
to $\varphi$, regardless of the structure. We use an image taken
immediately after homogenizing the sample to determine the
proportionality constant between $I$ and $\varphi$ and to correct
for any non-uniformity in the incident beam intensity profile. As
the gel sediments, the speckle pattern is shifted downwards: using
cross-correlation methods similar to those used in particle imaging
velocimetry~\cite{TokumaruExpInFluids1995}, we are able to measure
the full sedimentation velocity profiles, $v(z,t)$, with a vertical
resolution of about 0.5 mm. Finally, we
measure the microscopic dynamics as detailed in
Ref.~\cite{BrambillaPRL2011}, probing in particular particle motion
in the horizontal plane, on a length scale of the order of $q^{-1}
\approx R/1.5$.
%by calculating space- and time-resolved intensity correlation
%functions:
%\begin{equation}
%g_2(z,t,\tau)-1 = \frac{<I_p(t)I_p(t+\tau)>_z}{<I_p(t)>_z<I_p(t+\tau)>_z}-1\,,
%\label{eq:g2}
%\end{equation}
%where $<\cdot \cdot \cdot >_z$ is an average over all pixels
%belonging to a ROI at height $z$ and $I_p$ is the intensity of the
%$p$-th pixel. For a sedimenting sample, $I_p$ varies both because of
%the downward drift of the speckle pattern (unlike in usual dynamic
%light scattering) and because the particles change their relative
%position, i.e. due to the microscopic dynamics (like in usual
%dynamic light scattering). We use methods to be detailed in a
%forthcoming publication to separate the two contributions, so that
%our intensity correlation functions only reflect the relative motion
%of the particles. Moreover, owing to the orientation of
%$\mathbf{q}$, $g_2-1$ is only sensitive to motion in the horizontal
%plane, on a length scale of the order of $q^{-1} \approx R/1.5$,
%comparable to the particle size. We note that usually the
%depolarized scattered light is insensitive to collective dynamics,
%so that $g_2-1$ should in principle be proportional to the square of
%the self part of the intermediate scattering
%function~\cite{BrambillaPRL2011}. However, we expect structural
%rearrangements in a gel to involve the coordinated rotation of
%several particles, since rotational and translational motion will be
%coupled in a gel strand. Since depolarized light scattering is
%sensitive to particle rotation, $g_2-1$ is likely to contain
%contributions from both the self and the collective dynamics.
As a final remark, we emphasize that our measurements of $\varphi$,
of the microscopic dynamics and, to some extent, of $v$ all rely on
the assumption that the CCD images are formed only by singly
scattered photons. Fortunately, the relative low average refractive
index of the MFA particles ($n_p \approx 1.356$) allow one to
carefully match $n_p$ using water-based solvents, thereby
effectively suppressing multiple scattering.

For systems prepared from the second batch, we use a simplified
version of the apparatus shown in Fig.~\ref{fig:apparatus},
optimized for measuring only $h(t)$ and
$\varphi(z,t)$~\cite{Buzzaccaro}. It consists of a custom-made light
scattering setup, operating at a fixed scattering angle
$\theta=90^{\circ}$. Selection of the incident and detected
polarization of the scattered intensity is made by means of two
Glan-Thomson polarizers with an extinction ratio better than
$1\times10^{-6}$. The cell is mounted on a DC- motorized micrometric
translator allowing cell positioning with a resolution of $0.1 ~\mu
{\rm m}$ and an absolute accuracy of about $3 ~\mu {\rm m}$. A He-Ne
laser beam is mildly focused in the cell to a spot size $w = 46 ~\mu
{\rm m}$, corresponding to a depth of focus (Rayleigh range) of
about 10 mm, fixing the maximum useful optical path in the cell. The
whole setup is enclosed into a removable hood allowing to control
temperature to better than $0.5^{\circ}~C$.

\subsection{Visualization of stresses}
Valuable information on the distribution of the gravity-induced stress in the sample can be obtained by polarimetry. The sample is illuminated by a collimated beam of white light, issued from a LED source. A demagnified image of the gel is formed onto a CMOS camera, working in transmission. Two crossed polarizers are placed before and after the sample. Under these conditions, for a stress-free sample essentially no light is detected by the camera, because the birefringence induced by the particles is extremely weak. Because stress induces birefringence in the particles~\cite{MuellerJAP1935}, regions of the gel where gravity-induced stress accumulates depolarize the incident light and appear as bright in the image (see Fig.~\ref{fig:polarimetry} below for an example). More precisely~\cite{CokerCAMBRIDGE1930}, when the transmission axes of the polarizers are oriented at $\pm 45$ deg with respect to the vertical direction, stresses along the horizontal and vertical direction are visualized. Conversely, by turning the polarizers so that their axes are vertical and horizontal, stresses oriented at $\pm 45$ deg with respect to the vertical direction can be imaged.

\section{The poroelastic model}
\label{sec:poroelastic}

We shortly recall here the main features of the poroelastic model, a
popular model introduced by Biot~\cite{BiotJAP1941} and widely used in
the literature (see e.g.~\cite{BuscallFaradyTrans1987}) to describe the
deformation of a gel under its own weight. The system is treated as
a continuum medium that responds elastically to a compressive
deformation. The effect of the solvent back flow through the gel is
accounted for by introducing a viscous friction term. For the sake
of completeness, we introduce also a solid friction term, due to the
adhesion of the gel to the container walls, which is usually not
discussed in the literature.

Typically, the gel sedimentation occurs at a very low Reynolds
number and the rate of change of $v$ is small; thus, inertia terms can be neglected.
Using the reference frame shown in Fig.~\ref{fig:apparatus} with $z=0$ the cell bottom, Newton's law for a
gel slice of thickness $dz$ yields
\begin{equation}
\frac{\partial p}{\partial z}=-\Delta\rho\varphi
g-\frac{\partial\sigma}{\partial z}+\frac{\sigma}{L} \,.
\label{eq:poroelastic1}
\end{equation}
Here, $p$ is pressure, $\Delta \rho$ the buoyant density, $g$ the
acceleration of gravity, $\sigma$ the elastic stress, and $L$ a
characteristic length. The term on the l.h.s. originates from the
viscous drag, while the terms on the r.h.s. account for the buoyant
weight, the elastic stress and the wall friction, respectively. In
particular, $L$ represents Jansen's screening length
~\cite{DeGennesPRE1997,OvarlezPRE2003}, such that the
gravitational stress in a given horizontal plane $\Sigma$ exerted by
a gel slice located $\Delta z$ above $\Sigma$ is divided by a factor
$\exp(\Delta z L^{-1})$. This screening is due to the redirection of
part of the gel weight towards the walls. For $L$ to be finite, the
gel must ``push'' against the walls as it is compressed; in other
words, the gel must have a positive Poisson's ratio $\nu$, such that
$L = l(1-\nu)(\mu \nu)^{-1}$ is finite~\cite{DeGennesPRE1997,OvarlezPRE2003}, where $l$ is side of the square cell
section and $\mu$ is Coulomb's solid friction coefficient. We will
show in Sec.~\ref{sec:equilibriumprofiles} that for our gels $\nu
\approx 0$ and hence $L \rightarrow \infty$. Thus, in the following
we shall neglect the wall friction term in
Eq.~(\ref{eq:poroelastic1}).

The conservation of colloidal particles is expressed by the
continuity equation:
\begin{equation}
\frac{\partial\varphi}{\partial t}+\frac{\partial}{\partial
z}\left(v\varphi\right)=0 \, .
\label{eq:ContinuityGel}
\end{equation}
The particle velocity $v$ is related to the viscous stress gradient
by Darcy's law~\cite{darcy} via
\begin{equation}
q=-\frac{\kappa(\varphi)}{\eta}\frac{\partial p}{\partial z} \,,
\label{eq:Darcy}
\end{equation}
where $\kappa$ is the volume fraction dependent permeability, $\eta$
the solvent's viscosity, $q = (1-\varphi)(v_s-v)$ the fluid flux in
a reference frame co-moving with the gel, and $v_s$ the solvent
velocity in the laboratory frame. Since the volume of the suspension
(solvent plus particles) is conserved, $(1-\varphi)v_s=-\varphi v$,
which inserted in Eq.~(\ref{eq:Darcy}) yields
\begin{equation}
v=\frac{\kappa(\varphi)}{\eta}\frac{\partial p}{\partial z} \,.
\label{eq:Darcy2}
\end{equation}
By inserting Eq.~(\ref{eq:Darcy2}) in the continuity equation,
Eq.~(\ref{eq:ContinuityGel}), and replacing $\partial p / \partial
z$ by the r.h.s. of Eq.~(\ref{eq:poroelastic1}) with no solid
friction term, one finds
\begin{equation}
\frac{\partial\varphi}{\partial t}=\frac{\partial}{\partial
z}\left[\frac{\varphi\kappa(\varphi)}{\eta}\left(\Delta\rho\varphi
g+\frac{\partial\sigma}{\partial z}\right)\right] \,.
\label{eq:poroelastic2}
\end{equation}
The final equation describing the temporal evolution of the
concentration profiles in the poroelastic model is then
\begin{equation}
\frac{\partial\varphi}{\partial t}=\frac{\partial}{\partial
z}\left[\frac{\varphi\kappa(\varphi)}{\eta}\left(\Delta\rho\varphi
g+\frac{K(\varphi)}{\varphi}\frac{\partial\varphi}{\partial
z}\right)\right] \,,
\label{eq:poroelastic3}
\end{equation}
where we have used the definition of the uniaxial compressional modulus $K$
\begin{equation}
K(\varphi)=\varphi\frac{\partial\sigma}{\partial\varphi}\,.
\label{eq:Kphi}
\end{equation}

In order to solve Eq.~(\ref{eq:poroelastic3}), the two material
functions $\kappa(\varphi)$ and $K(\varphi)$ must be specified.
Insight on $K(\varphi)$ may be gained by examining the gel behavior
for $t\rightarrow \infty$, when the gels approaches mechanical
equilibrium. In this limit, $v$ vanishes and so does the viscous
friction. If wall friction is negligible, at any height $z$ the
buoyant weight per unit area of the gel column above $z$ is balanced
by the elastic response of the gel portion below $z$:
\begin{equation}
\int_{z}^{h_{\infty}} \Delta \rho g \varphi {\rm d}z = \sigma(z) \,,
\label{eq:asymptoticeq}
\end{equation}
implying that $\varphi$ must satisfy
\begin{equation}
\Delta \rho g \varphi = -\frac{\partial \sigma}{\partial z} =-\frac{\partial \sigma}{\partial \varphi}\frac{\partial \varphi}{\partial z}\,.
\label{eq:partialphieq}
\end{equation}
Experimentally, Eq.~(\ref{eq:asymptoticeq}) can be used to obtain
$\sigma(z)$, provided that $\varphi(z)$ can be measured. Once both
$\sigma(z)$ and $\varphi(z)$ are known, the volume fraction
dependence of the elastic stress is simply obtained by plotting
directly $\sigma$ \textit{vs} $\varphi$. As we will show in
Sec.~\ref{sec:equilibriumprofiles}, for our gels $\sigma(\varphi)$
is well modeled by a power law:
\begin{equation}
\sigma(\varphi)=A\varphi^{\alpha} \,.
\label{eq:sigmavsphi}
\end{equation}
Assuming this functional form, Eq.~(\ref{eq:partialphieq}) can be easily solved:
\begin{equation}
\varphi(z) = \left[  \frac{\Delta \rho g}{A}\frac{\alpha-1}{\alpha}(z_{\rm max}-z) \right ] ^{\frac{1}{\alpha-1}} \,,
\label{eq:phieq}
\end{equation}
where the integration constant $z_{\rm max}$ is determined by imposing particle conservation, finding
\begin{equation}
z_{\rm max}^{\frac{^\alpha}{\alpha-1}} - \left ( z_{\rm max} - h_{\infty} \right) ^{\frac{\alpha}{\alpha-1}} = h_0\varphi_0 \left ( \frac{\alpha}{\alpha-1}    \right) ^{\frac{\alpha}{\alpha-1}}  \left ( \frac{A}{\Delta \rho g} \right ) ^ { \frac{1}{\alpha-1}}\,.
\label{eq:C}
\end{equation}
%Note that the above procedure is justified only if one assumes that
%the static modulus $K(\varphi)$ measured in the limit $v \rightarrow
%0$ coincides with the the modulus at finite strain rate that is
%relevant during sedimentation.

The determination of the permeability $\kappa(\varphi)$ is less
straightforward. In practice, we test several functional forms
proposed for $\kappa(\varphi)$ and check whether they are able to
reproduce the temporal evolution of $h(t)$, the total gel height, as we
shall discuss it in Sec.~\ref{sec:kynetics}.
\begin{equation}
\label{eq:}
\end{equation}

\section{Results and discussion}

\subsection{Equilibrium concentration profiles}
\label{sec:equilibriumprofiles}
\begin{figure}
\begin{center}
\includegraphics[width = 8 cm]{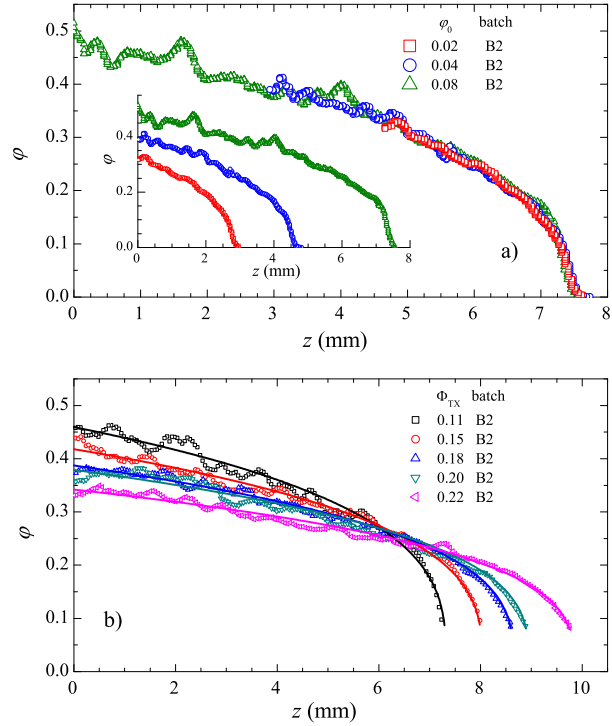}
\end{center}
\caption{Inset of a): asymptotic concentration profiles for gels
with fixed $\phi_{\rm TX} = 0.12$ and various particle
concentrations, as indicated in the labels. Main plot of a): same
data rescaled onto a master curve as described in the text. b):
asymptotic concentration profiles for gels with fixed $\varphi_0=0.08$
and various surfactant concentrations, as indicated by the labels.
Lines: fits of Eq.~(\ref{eq:phieq}) to the data.}
\label{fig:profiles}
\end{figure}

In this section, we focus on the particle volume fraction and
interaction strength dependence of the concentration profiles in the
long time limit, when no sedimentation is experimentally detectable
and the gel is very close to mechanical equilibrium.

For gels of batch B1 and B2, concentration profiles where measured
at least 240 and 500 hours after initializing
the sample, respectively. We first focus on gels of batch B2.
Figure~\ref{fig:profiles}a) shows $\varphi_{\infty}(z) =
\varphi(z,t\rightarrow \infty)$ for three gels, prepared at various
particle concentrations $\varphi_0$ but keeping the same particle
interactions ($\phi_{\rm TX} = 0.12$). The main plot shows that the
shape of the concentration profiles does not depend on $\varphi_0$:
the profiles can be collapsed onto a master curve simply by shifting
them along the $z$ axis so as to make the top part coincide.
%by an amount $z_{\rm max} \approx
%h_{\infty}$. 
Thus, the final height of a gel with a given
$\varphi_0$ may be simply predicted using the master curve of
Fig.~\ref{fig:profiles} a) and imposing mass conservation. Note that
this surprising scaling implies that gels prepared at different
$\varphi_0$ compress in the same way and thus must have the same
strength, a counterintuitive result. In fact, we anticipate that
these gels are formed by the debris of a percolating network, which
is too weak to sustain itself and collapses almost immediately.
These debris must have similar structure regardless of $\varphi_0$,
so that the network resulting from their accumulation at the bottom
of the cell has essentially the same mechanical properties.

Interestingly, we observe that the concentration profiles are not
very smooth. Measurements of $\varphi_{\infty}(z)$ for the same gels
but using the ``beam deflection'' method described in
Ref.~\cite{PiazzaJPCM2011} exhibit a smoother behavior, probably
because in the beam deflection method the concentration profile is
smoothed over a length comparable to the cell optical path
($5~\mathrm{mm}$). The noise in $\varphi$ as measured here may be
due also to an extra-contribution to the intensity of the
depolarized scattered light stemming from stress-induced
birefringence, which we will show below to be spatially non-uniform.
In spite of these differences, we emphasize that there is an overall
good agreement between $\varphi$ measured by the beam deflection and
the depolarized scattered intensity methods~\cite{PiazzaJPCM2011}.

The concentration profiles of gels prepared at a fixed particle
concentration $\varphi_0 = 8 \%$ and variable interaction strength
are shown in Fig.~\ref{fig:profiles} b). As $\phi_{\rm TX}$
increases, interparticle attractive forces grow and the gel becomes
stronger, leading to a smaller compaction under the action of
gravity. The change in gel strength is reflected by the fact that
the shape of the asymptotic profiles is modified. Indeed, in this
case the simple scaling shown in Fig.~\ref{fig:profiles} a) is no
more possible.

\begin{figure}
\begin{center}
\includegraphics[width = 8 cm]{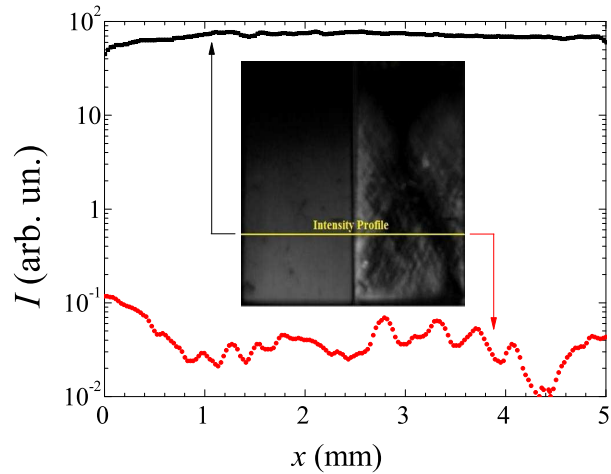}
\end{center}
\caption{Inset: visualization of gravity-induced stress in a gel of
batch B2 observed between crossed polarizers ($\varphi = 0.08,
\Phi_{\rm TX}=0.12$). The height of the image corresponds to the gel
height, the width of the cell is 5 mm. In the image on the left
(resp., on the right) the polarizers are oriented so as to visualize
stress along the vertical and horizontal directions (resp., along
directions at $\pm 45$ deg with respect to vertical). The right
image was obtained using an exposure time 1000
times larger than that for the left image. Main plot: intensity
level along the horizontal line shown in the images. The intensity
levels have been corrected for the difference in exposure time.}
\label{fig:polarimetry}
\end{figure}

As discussed in Sec.~\ref{sec:poroelastic}, the asymptotic
concentration profile is dictated by the $\varphi$ dependence of the
elastic stress $\sigma$, which may be obtained directly from the
experimental $\varphi_{\infty}(z)$, provided that wall friction is
negligible. To investigate whether this assumption holds for our
gels, we image the distribution of stresses in the gel by
polarimetry. The inset of Fig.~\ref{fig:polarimetry} shows the gel
column observed under crossed polarizers, where highly stressed
regions are brighter. In the left image, stresses along the vertical
and horizontal direction are \emph{a priori} imaged, although one of
course expects vertical stress to dominate. The brightness increases
towards the bottom of the cell, a first indication that most of the
gravitational stress is transmitted downward, rather than being
redirected to the walls. In the right image, the crossed polarizers
have been rotated, in order to visualize stress oriented at $\pm 45$
deg with respect to the vertical direction, which would be
non-negligible if wall friction was significant. Although some
bright patches do appear, we emphasize that the right image was
taken with an exposure time 1000 times larger than for the left one.
The intensity profiles along an horizontal line, corrected for the
exposure time, clearly show that the vertical component of the
stress dominates over that at $\pm 45$ deg, implying that wall
friction can be neglected for our gels. Interestingly, this is at
variance with observations for other kinds of colloidal
gels~\cite{CondreJSTAT2007} and may be due to the fact that here the
percolating network initially formed is then broken, in contrast to
the gels of Ref.~\cite{CondreJSTAT2007}.

\begin{figure}
\begin{center}
\includegraphics[width = 8 cm]{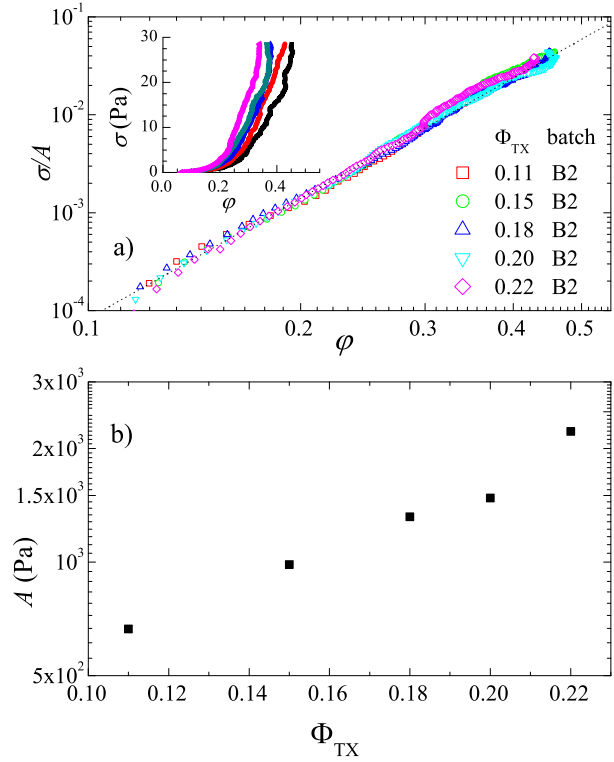}
\end{center}
\caption{Inset of a): elastic stress as a function of $\varphi$, for
gels with $\varphi_0 = 0.08$ and $\Phi_{\rm TX}$ as indicated by the
labels. Main plot: scaling of $\sigma(\varphi)/A$ onto a single
power law. The dotted line has slope 4.08. b) Dependence of the
scaling factor $A$ on the amount of depletant.} \label{fig:sigmavsphi}
\end{figure}

From the asymptotic concentration profiles shown in
Fig.~\ref{fig:profiles} b), we calculate $\sigma(\varphi)$ as
described in Sec.~\ref{sec:poroelastic} (see
Eq.~(\ref{eq:asymptoticeq})) and plot the result in the inset of
Fig.~\ref{fig:sigmavsphi} a). The main graph shows that data for all
$\Phi_{\rm TX}$ can be collapsed onto a single line in a log-log
plot, thus indicating that the stress grows with $\varphi$ as a
power law: $\sigma = A\varphi^{\alpha}$. The exponent is essentially
the same for all curves, as shown by the good collapse in
Fig.~\ref{fig:sigmavsphi}. By contrast, the prefactor $A$ depends on
$\Phi_{\rm TX}$: stronger bonds (larger $\Phi_{\rm TX}$) yield a
stiffer gel. One expects $A$ to scale with the interparticle bond
spring constant, $A \sim U_c/r^2$, where $U_c$ is the interparticle
potential at contact. If the Triton micelles were non-interacting,
$U_c$ and thus $A$ should scale with $\Phi_{\rm
TX}$~\cite{AsakuraJChemPhys1954}, whereas Fig.~\ref{fig:sigmavsphi}
b) rather suggests $A \sim \exp [\Phi_{\rm TX}]$. Thus, this
behavior hints at deviations with respect to the ideal depletion
interaction induced by a diluted gas of micelles.

In principle, both $A$ and $\alpha$ may be obtained by fitting
$\sigma(\varphi)$ to a power law. However, we find that a more
robust procedure consist in fitting the concentration profiles
$\varphi(z,t \rightarrow \infty)$ to Eq.~(\ref{eq:phieq}), the
expression derived in Sec.~\ref{sec:poroelastic} for a power law
dependence of $\sigma$ \textit{vs} $\varphi$. The resulting fits are
shown as lines in Fig.~\ref{fig:profiles} b). We find that all the
concentration profiles for the gels of Fig.~\ref{fig:profiles} b)
can be very well fit using the same value $\alpha = 4.08$. We
furthermore check that the fitting parameter $z_{\rm max}$ of
Eq.~(\ref{eq:phieq}) satisfies Eq.~(\ref{eq:C}) to within $2\%$. It
is worth noting that, although $\sigma$ must eventually diverge as
$\varphi$ approaches the volume fraction of random close packing, no
hint of such divergence is observed here, nor in
Ref.~\cite{BrambillaPRL2011}, up to $\varphi$ values as high as 0.45
(batch B2) or even 0.55 (batch B1).

\begin{figure}
\begin{center}
\includegraphics[width = 8 cm]{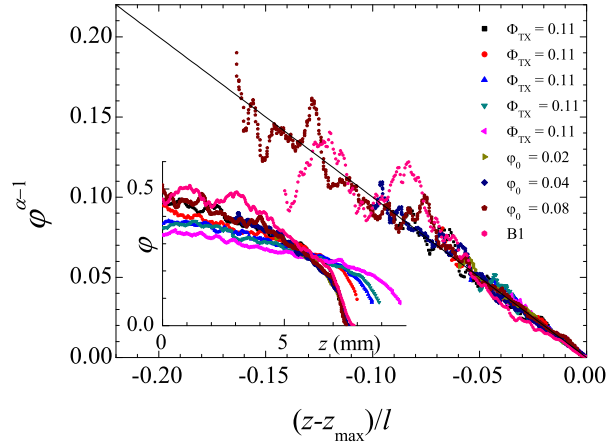}
\end{center}
\caption{Inset: asymptotic concentration profiles for all gels of
both batches. Main plot: the same data collapse onto a straight line
when using reduced variables, as explained in the text. Data for
gels of batch B1 at $\varphi_0 = 0.08$ (resp., $\Phi_{\rm TX} =
0.12$) are labeled by $\Phi_{\rm TX}$ (resp., $\varphi_0$). For the
gel of batch B1, $\varphi_0 = 0.12$, $\phi_{\rm TX} = 0.12$. The
line is the poroelastic model.} \label{fig:superscaling}
\end{figure}

We check that the law $\sigma = A\varphi^{\alpha}$ holds for all
gels of both batches, with similar exponents: $\alpha = 3.6$ (resp., 4.1)
for batch B1 (resp., B2). Given the power law dependence of
 $\sigma$ \textit{vs} $\varphi$, Eq.~(\ref{eq:phieq}) suggests that all concentration profiles should
collapse on the same straight line when plotting
$\varphi^{\alpha-1}$ \textit{vs} $(z-z_{\rm max})/\ell$, where $\ell
= A/(\Delta \rho g)$ is a characteristic length scale that compares
the gel elasticity to the gravity pull. Figure
\ref{fig:superscaling} shows that this is indeed the case for all
the gels we have studied.

\subsection{Sedimentation kinetics}
\label{sec:kynetics}

\begin{figure}
\begin{center}
\includegraphics[width = 8 cm]{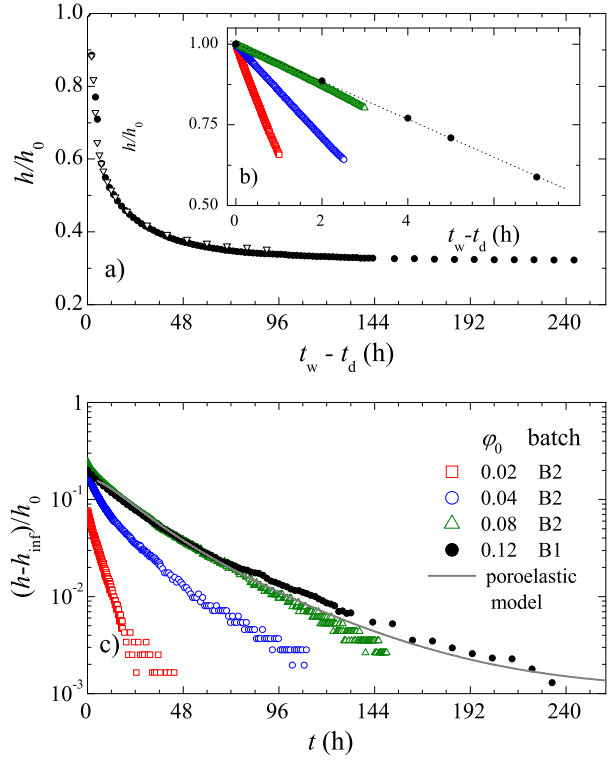}
\end{center}
\caption{a): time evolution of the total height for a gel of batch
B1 with $\varphi = 0.12$, $\phi_{\rm TX} = 0.12$. The cell section
is $3\times 3~{\rm mm}^2$ ($5\times 5~{\rm mm}^2$) for the solid
circles (open triangles), $h_0 = 22.7$ mm. b): initial decay of $h$
during the collapse of gels of both batches (same symbols as in c)):
in all cases, a linear behavior is observed. c) semilogarithmic plot
of the time varying part of $h$ normalized by $h_0$, during the
compaction regime. The line is a fit of the poroelastic model to the
data for the gel of batch B1.} \label{fig:hoft}
\end{figure}

The temporal evolution of the total height of the gel, $h$, exhibits
a qualitatively similar behavior for all the gels of batches B1 and
B2. Denoting by $t_\mathrm{w}$ the time after filling the sample
cell, an initial regime where no sedimentation occurs is observed
for $t_\mathrm{w} < t_\mathrm{d}$. The latency time $t_d$ grows with
the strength of interparticle bonds and $\varphi_0$, as observed in
previous
works~\cite{StarrsJPCM2002,kilfoil03,huh07,buscall09,BartlettArxiv}.
During this phase, the speckle dynamics is essentially frozen,
indicating that a system-spanning arrested network has been formed.
For $t_\mathrm{d} \leq t_\mathrm{w} < t_\mathrm{c}$, $h$ decreases
steeply and the speckles fluctuate rapidly, indicating that the
network has failed under gravitational stress. Finally, for
$t_\mathrm{w} \geq t_\mathrm{c}$ the falling debris of the initial
network have deposited at the bottom of the cell, forming a denser
gel, as indicated by the dramatic slowing down of the speckle
dynamics. This denser gel slowly compacts under its own weight: we
define $t = t_\mathrm{w}-t_\mathrm{c}$ and describe the gel
evolution in this third regime using the poroelastic model of
Sec.~\ref{sec:poroelastic}. An example of $h(t_\mathrm{w})$ for a
gel of batch B1 is shown in Fig.~\ref{fig:hoft} a). Here, network
failure occurs almost immediately, $t_\mathrm{d}\approx 0$. Note
that $h(t_\mathrm{w})$ is almost identical for gels prepared in
cells with square sections equal to $3 \times 3 ~{\rm mm}^2$ and $5
\times 5 ~{\rm mm}^2$, thus confirming that wall effects are
negligible. Interestingly, during the initial collapse
$h(t_\mathrm{w})$ decreases linearly in time, as observed also for
gels of the second batch prepared at various $\varphi_0$, see
Fig.~\ref{fig:hoft} b). This behavior is in striking contrast with
that reported very recently by Bartlett \textit{et
al.}~\cite{BartlettArxiv}, who find a ``compressed exponential''
decrease, $h(t_\mathrm{w}-t_\mathrm{d}) \sim
\exp\{-[(t_\mathrm{w}-t_\mathrm{d})/\tau_c] ^{1.5}\}$. This
difference might be due to the different range of the attractive
potential, which is small in our case ($r/R \approx 0.037$) and much
larger in Ref.~\cite{BartlettArxiv} ($r/R \approx 0.62$). More
experiments will be needed to elucidate this point.
Figure~\ref{fig:hoft} c) shows the time-varying part of $h$ during
the compaction stage, for various $\varphi_0$. The decay of $h$ is
not a simple exponential, as shown by the curvature of the data in a
semilogarithmic plot. Indeed, we find that the data can be
reasonably well fit by a stretched exponential function,
$h(t)-h_{\infty} \sim \exp[-(\Gamma t)^{\beta}]$, with $\beta =
0.68-0.87$, depending on sample composition (fit not shown).
Although the stretched exponential fit works well, it has no obvious
theoretical interpretation, besides the generic remark that the
system exhibits a distribution of relaxation times. A better
justified fit is provided by the prediction of the poroelastic
model, Eq.~(\ref{eq:poroelastic3}), which we solve numerically for
the gel of batch B1~\cite{BrambillaPRL2011} and show as a continuous
line in Figure~\ref{fig:hoft} c). As discussed in
Sec.~\ref{sec:poroelastic}, the  permeability $\kappa(\varphi)$ must
be provided in order to solve Eq.~(\ref{eq:poroelastic3}). By
testing various functional forms, including those proposed in the
past for gels with a fractal structure~\cite{KimPRL2007}, we find
that only a critical-like law similar to that used for suspensions
of hard spheres is able to reproduce our results: $\kappa(\varphi) =
\kappa_0 \varphi^{-1}(1-\varphi)^m$, with $m$ = 7. The remarkable
success of the poroelastic model is demonstrated by the fact that,
having determined $K(\varphi)$ and $\kappa(\varphi)$ from fits of
$\varphi(z,t\rightarrow \infty)$ and $h(t)$, respectively, the
solution to Eq.~(\ref{eq:poroelastic3}) reproduces very well the
full time evolution of the concentration and velocity profiles, as
reported in Ref.~\cite{BrambillaPRL2011}.

\begin{figure}
\begin{center}
\includegraphics[width = 8 cm]{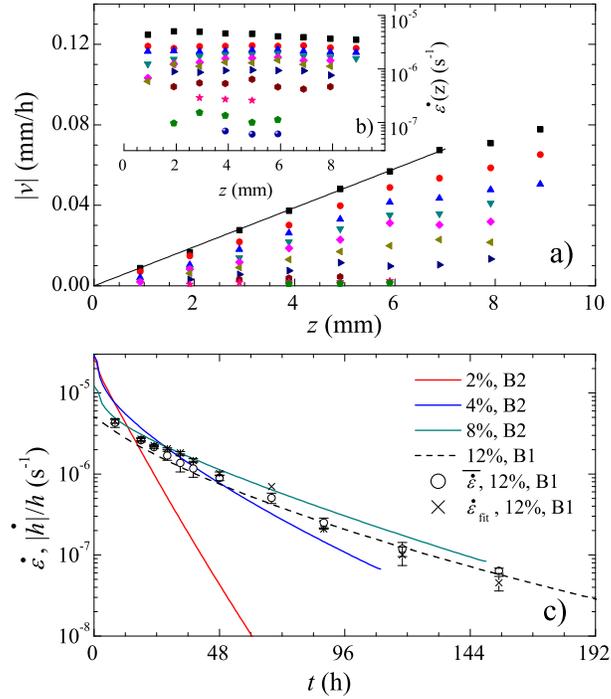}
\end{center}
\caption{a) sedimentation velocity \textit{vs} height at various
times, ranging from $t_{\rm w} = 30$h to $t_{\rm w} = 167$h from top
to bottom, for a B1 gel with $\varphi = 0.12$, $\Phi_{\rm TX} =
0.12$. The line is a linear fit to $v(z,t_{\rm w}=30\mathrm{h})$ up
to $z=7$ mm. b): strain measured over a height $z$,
$\dot{\varepsilon}=|v|/z$, for the same data as in a). c): global
strain $|\dot{h}|/h$ (dotted line), $z-$averaged strain
$\overline{\dot{\varepsilon}}$ (open circles) and
$\dot{\varepsilon}_{\mathrm{fit}}$ as defined in the text, for the
same gel as in a) and b). The solid lines are $|\dot{h}|/h$ for the
three gels of batch B2 shown in Fig.~\ref{fig:hoft}, for which
$\Phi_{\rm TX}$ increases from left to right.} \label{fig:epsdot}
\end{figure}

One of the most surprising findings of Ref.~\cite{BrambillaPRL2011}
is the linear variation of the sedimentation velocity $v$ with
height $z$ in the gel column. Figure~\ref{fig:epsdot} a) shows that
indeed $|v| \sim  z$, at least for the lower part of the gel column.
Close to the top, deviations from this linear behavior appear, in
that $v$ appears to grow more slowly with $z$. These deviations are
more pronounced for longer  times. We define a compressive strain
rate, $\dot{\varepsilon}_{\rm fit}$, as the slope of a linear fit to
$v(z)$ in the regime where a linear behavior is observed, e.g. up to
$z=7$ mm for data at $t_{\rm w} = 30$h in Fig.~\ref{fig:epsdot} a).
We also define the strain rate measured at height $z$ as
$\dot{\varepsilon}(z) = v/z$ and plot its height dependence in
Fig.~\ref{fig:epsdot} b). This quantity is almost independent of
$z$, with the exception of the very bottom and top parts of the gel
column, thus confirming that the temporal evolution of the full
velocity profiles is essentially captured by the evolution of a
single parameter, $\dot{\varepsilon}$. Additionally, we have shown
in Ref.~\cite{BrambillaPRL2011} that $\dot{\varepsilon}$ governs
also the microscopic dynamics, since structural rearrangements occur
on a time scale $\tau_{\alpha}$ such that
$\tau_{\alpha}\dot{\varepsilon} = \varepsilon_\mathrm{y}$, where
$\varepsilon_\mathrm{y}\approx 0.02$ is the typical yield strain
beyond which irreversible rearrangements occur. In view of the key
role played by $\dot{\varepsilon}$, it is interesting to test
whether a macroscopic measurement of the global strain rate,
$|\dot{h}|/h$,  reflects accurately the microscopic strain rate. In
Fig.~\ref{fig:epsdot} c) we compare for a gel of batch B1
$|\dot{h}|/h$ (dotted line) to $\dot{\varepsilon}_{\rm fit}$
(crosses) and $\overline{\dot{\varepsilon}}$, the average of
$\dot{\varepsilon}(z)$ over the gel height (open circles). In order
to reduce the impact of the measurement noise on the numerical
derivative of $h(t)$, we  differentiate the stretched exponential
fit of $h(t)$ discussed in reference to Figure~\ref{fig:hoft} c),
rather than the data themselves. Figure~\ref{fig:epsdot} c) shows
that a reasonably good agreement is found between these three ways
of quantifying the compressive strain rate, thus demonstrating that
deviations from the linear behavior of $v(z)$ are limited. To check
the generality of our findings we simulate the time evolution of
gels with different value of $K(\varphi)$ and $\kappa(\varphi)$. In
all cases we have studied, $|v| \sim  z$ in the lower part of the
gel column while, close to the top, $v$ grows slower with $z$.
Besides $\dot{\varepsilon}_{\rm fit}$, $\dot{\varepsilon}(z)$ and
$|\dot{h}|/h$ are strongly correlated. In particular, we found that
$|\dot{h}|/h \, < \, \dot{\varepsilon}(z) = v/z\, <\,
\dot{\varepsilon}_{\rm fit}\,\approx \,A|\dot{h}|/h$ with $1<A<1.35$
depending on the functional form of $K(\varphi)$ and
$\kappa(\varphi)$, so that  $\dot{\varepsilon}$ may be directly
estimated from $|\dot{h}|/h$. This observation can be particularly
useful when dealing with turbid samples, where the velocity profiles
can not be measured directly.

Figure~\ref{fig:epsdot} c) shows also the macroscopic strain rate
for the same three gels of batch B2 as in Figs.~\ref{fig:profiles}
a) and~\ref{fig:hoft}. In spite of the variation of $\varphi_0$, the
strain rate at the beginning of the compaction stage is of the same
order of magnitude for all samples. This is probably due to the fact
that the structure of the gel is similar, as discussed in relation
to the scaling of the concentration profiles shown in
Fig.~\ref{fig:profiles} a). By contrast, the temporal evolution of
the strain rate is significantly different, $|\dot{h}|/h$ decaying
faster for the gels at lower initial $\varphi$. This is at odd with
the scaling behavior of the concentration profiles reported in
Fig.~\ref{fig:profiles}a, where it was shown that the top part of
the gels at fixed $\Phi_{\rm TX}$ have the same concentration
profile. This can be qualitatively understood be recalling that, although the top of the profiles for
samples prepared at different $\varphi_0$ are similar (see
Fig.\ref{sec:equilibriumprofiles}), gels prepared at higher initial
concentration reach a higher volume fraction at the cell bottom.  Therefore,
they display in this region a lower permeability and a higher stiffness, which slows down the
compaction process.

%\textbf{OLD:Qualitatively, the $\varphi_0$ dependence of
%$|\dot{h}|/h$ can be understood by noticing that the solvent back
%flow is faster for the taller gels, i.e. those prepared at a larger
%$\varphi_0$. This stronger back flow counteracts more effectively
%the gravitational stress, leading to a slower compaction process}.
%This behavior can be understood by considering that the compaction
%is more pronounced at the bottom of the sample, where the local
%strain rate is maximum, and $h(t)$ is essentially governed by the
%dynamics of this region. 

\section{Conclusions}
\label{sec:conclusions}

We have studied the sedimentation kinetics and the concentration profiles of depletion-induced colloidal gels, varying both the particle volume fraction and the strength of the interparticle interactions. By using optical methods that combine light scattering and imaging, we have been able to gain detailed information on the evolution of the concentration profiles and the dynamics of the gels. Coupled to stress visualization experiments that rule out any significant role of solid friction on the cell walls, this has allowed us to measure quantitatively the volume fraction dependence of the elastic response of the gels and to test thoroughly the poroelastic model. In discussing the $\varphi_0$ and $\Phi_{\rm TX}$ dependence of the parameters issued from the poroelastic model, it is essential to keep in mind that the model can only be applied to the last regime of the sedimentation process, i.e. the compaction of a denser gel formed by the falling debris of the initial network that fails under gravity. This allows one to rationalize some apparently paradoxical results, such as the scaling of the asymptotic concentration profiles obtained at fixed $\Phi_{\rm TX}$ and variable $\varphi_0$. 

Our work and that of other groups using confocal microscopy~\cite{huh07,BartlettArxiv} show that detailed microscopic information on the structure and the dynamics of the gels can be very valuable in order to better understand the behavior of colloidal gels under gravitational stress. More work along these lines will be required to fully understand the microscopic origin of the diversity of behaviors observed for different systems.

\ack We thank L. Berthier for many useful discussions and
Solvay-Solexis for having provided us the Hyflon$^{TM}$ MFA particle
batch. Financial support from R\'{e}gion Languedoc Roussillon, CNES,
ANR ``Dynhet'', the Italian Ministry of Education and Research (MIUR)
(PRIN 2008) and ASI is gratefully acknowledged.

\section*{References}
\bibliographystyle{unsrt}
%\bibliography{biblio_gel}

\begin{thebibliography}{}

\end{thebibliography}


\begin{thebibliography}{10}

\bibitem{ZaccarelliJPCM2007}
Zaccarelli E.
\newblock {\em J. Phys.:Condens. Matter}, 19(32):323101, 2007.

\bibitem{Buzzaccaro}
Buzzaccaro S, Rusconi R, and Piazza R.
\newblock {\em Phys. Rev. Lett.}, 99:098301, 2007.

\bibitem{LuNature2008}
Lu~PJ, Zaccarelli E, Ciulla F, Schofield A, Sciortino F, and Weitz
DA.
\newblock {\em Nature}, 453(7194):499--503, 2008.

\bibitem{StarrsJPCM2002}
Starrs L, Poon WCK, Hibberd DJ, and Robins MM.
\newblock {\em J. Phys.:Condens. Matter}, 14(10):2485--2505, 2002.

\bibitem{kilfoil03}
Kilfoil ML, Pashkovski EE, Masters JA, and Weitz DA.
\newblock {\em Philos. Trans. R. Soc. A-Math. Phys. Eng. Sci.},
  361(1805):753--766, 2003.

\bibitem{huh07}
Huh JY, Lynch ML, and Furst EM.
\newblock {\em Phys. Rev. E}, 76, 2007.

\bibitem{buscall09}
Buscall R, Choudhury TH, Faers MA, Goodwin JW, Luckham PA, and
Partridge SJ.
\newblock {\em Soft Matter}, 5(7):1345--1349, 2009.

\bibitem{BartlettArxiv}
Bartlett P, Teece LJ, and Faers MA.
\newblock {\em arxiv}, page 1109.4893, 2011.

\bibitem{AllainPRL1995}
Allain C, Cloitre M, and Wafra M.
\newblock {\em Phys. Rev. Lett.}, 74(8):1478--1481, 1995.

\bibitem{CondreJSTAT2007}
Condre JM, Ligoure C, and Cipelletti L.
\newblock {\em J Stat Mech - Theory and Experiments}, page P02010, 2007.

\bibitem{ManleyPRL2005Sedimentation}
Manley S, Skotheim JM, Mahadevan L, and Weitz DA.
\newblock {\em Phys. Rev. Lett.}, 94(21):218302, 2005.

\bibitem{KimPRL2007}
Kim C, Liu Y, Kuhnle A, Hess S, Viereck S, Danner T, Mahadevan L,
and Weitz DA.
\newblock {\em Phys. Rev. Lett.}, 99(2):028303, 2007.

\bibitem{DerecPRE2003}
Derec C, Senis D, Talini L, and Allain C.
\newblock {\em Phys. Rev. E}, 67(6):062401, 2003.

\bibitem{BiotJAP1941}
Biot MA.
\newblock {\em J. Appl. Phys.}, 12:155--164, 1941.

\bibitem{BuscallFaradyTrans1987}
Buscall R and White LR.
\newblock {\em J. Chem. Soc., Faraday Trans.}, 83:873--891, 1987.

\bibitem{BrambillaPRL2011}
Brambilla G, Buzzaccaro S, Piazza R, Berthier L, and Cipelletti L.
\newblock {\em Phys. Rev. Lett.}, 106:118302, 2011.

\bibitem{AsakuraJChemPhys1954}
Asakura S and Oosawa F.
\newblock {\em J. Chem. Phys.}, 22:1255--1256, 1954.

\bibitem{BaxterJChPh1968}
Baxter RJ.
\newblock {\em J. Chem. Phys.}, 49:2770--2774, 1968.

\bibitem{DuriPRL2009}
Duri A, Sessoms DA, Trappe V, and Cipelletti L.
\newblock {\em Phys. Rev. Lett.}, 102(8):085702--4, 2009.

\bibitem{buzzaccaro3}
Buzzaccaro S, Piazza R, Colombo J, and Parola A.
\newblock {\em J. Chem. Phys.}, 132(12):124902, 2010.

\bibitem{TokumaruExpInFluids1995}
Tokumaru PT and Dimotakis PE.
\newblock {\em Exp. Fluids}, 19:1--15, 1995.

\bibitem{MuellerJAP1935}
Mueller H.
\newblock {\em J. Appl. Phys.}, 6:179--184, 1935.

\bibitem{CokerCAMBRIDGE1930}
Coker EG and Filon LNG.
\newblock {\em Treatise on Photoelasticity}.
\newblock Cambridge Press, 1930.

\bibitem{DeGennesPRE1997}
Boutreux T, Rapha\"{e}l E, and de~Gennes~PG.
\newblock {\em Phys. Rev. E}, 55(5):5759--5773, 1997.

\bibitem{OvarlezPRE2003}
Ovarlez G, Fond C, and Cl\'ement E.
\newblock {\em Phys. Rev. E}, 67(6):060302, 2003.

\bibitem{darcy}
Darcy H.
\newblock {\em Les Fontaines Publiques de la Ville de Dijon}.
\newblock Dalmont, Paris, 1856.

\bibitem{PiazzaJPCM2011}
Piazza R, Buzzaccaro S, and Secchi E.
\newblock {\em J. Phys.: Condens. Matter}, this issue, 2011.

\end{thebibliography}

\end{document}